\documentclass[letterpaper,10pt]{article}
\usepackage{osameet2}

\usepackage{amsmath,amssymb}

\usepackage{graphicx}
\usepackage[font=small]{caption}
\usepackage[font=footnotesize]{subcaption}

\usepackage[pdftex,colorlinks=true,bookmarks=false,citecolor=blue,urlcolor=blue]{hyperref} %
\usepackage{epstopdf}
\usepackage{booktabs}
\usepackage{color}

\newcommand{\EbNo}{E_{\text{b}}/N_{\text{0}}}

\newcommand{\w}{w}
\newcommand{\BB}{\mathsf{B}}

\begin{document}

\title{Energy-Efficient Soft-Assisted Product Decoders}

\vspace{-0.1cm}

\author{Christoffer Fougstedt$^1$, Alireza Sheikh$^2$, Alexandre Graell i Amat$^2$, Gianluigi Liva$^3$,\\ and Per Larsson-Edefors$^1$}
\address{$^1$Dept.\ of Computer Science and Engineering,
$^2$Dept.\ of Electrical Engineering,
Chalmers University of Technology, Sweden\\
$^3$Institute of Communications and Navigation, German Aerospace Center (DLR), Germany}
\email{chrfou@chalmers.se}

\vspace{-0.1cm}

\begin{abstract}
We implement a 1-Tb/s 0.63-pJ/bit soft-assisted product decoder in a 28-nm technology. The decoder uses one bit of soft information to improve its net coding gain by 0.2\,dB, reaching 10.3--10.4 dB, which is similar to that of more complex hard-decision staircase decoders.
\end{abstract}
\vspace{1.5 pt}

\section{Introduction}

\label{sec:intro}

Forward error correction (FEC) codes decoded using
hard-decision (HD) decoding, such as
product and staircase
codes, are commonly considered for moderate-reach
high-throughput fiber-optic communication systems as they offer
a relatively high net coding gain (NCG) while using
low-complexity decoders. But often there is soft information
available in DSP-based fiber-optic receivers; soft information
that can be harnessed to improve decoding performance. In
contrast to soft-decision (SD) decoding, which typically
operates using fixed-point math and entails a significant cost
in implementation complexity for operations, data storage, and
update requirements, hybrid HD-SD decoding schemes, such as
binary message-passing low-density parity-check (LDPC) decoders operating on quantized
channel information~\cite{lechner+:tcom12} and recently
proposed low-complexity decoding algorithms for product
codes~\cite{sheikh+:ECOC18,sheikh+:ISTC18}, can allow a decoder
to operate internally on hard decisions, but selectively use
soft information during decoding to significantly improve the
NCG. With the rationale that the NCG can be improved at a very
minor increase in implementation complexity over HD decoders,
we propose and evaluate a VLSI architecture of a {\it
soft-assisted product decoder} based on the iBDD-SR
algorithm~\cite{sheikh+:ECOC18,sheikh+:ISTC18}, with a decoder
core based on our earlier staircase decoder
architecture~\cite{fougstedt+:OFC2018}.

The proposed soft-assisted product decoder uses a single bit of
soft information. Remarkably, this allows us to achieve similar
coding gains as more complex staircase
decoders~\cite{zhang+:JLT14,fougstedt+:OFC2018}. While
staircase decoders provide high NCG using pure HD decoding,
they require a large data-processing memory due to the windowed
decoding operation. In contrast, the proposed iBDD-SR
soft-assisted product decoder has lower circuit
complexity, leading to less circuit area and decoding latency. %
In addition, since the decoder core uses hard decisions only,
it can support significantly higher throughputs than SD
decoders~\cite{larsson-edefors+:SPPCOM18}, making our proposed
iBDD-SR soft-assisted product decoder a viable option for
very-high-throughput high-NCG systems.

\section{Decoder Algorithm and Architecture}

Product codes can provide high NCG when employing
low-complexity iterative bounded-distance decoding
(iBDD)~\cite{justesen+:TCOM2011}. In iBDD, bounded
distance decoding of the component codes is
performed in an iterative row/column manner. However, there is a
probability that the component decoders introduce
miscorrections, which reduce the achievable NCG. Recently,
iBDD-SR was proposed~\cite{sheikh+:ECOC18}, %
in which the channel soft information is used to reduce the
component-decoder miscorrection rate. %
Here, we briefly review the iBDD-SR algorithm and show that we
can simplify the algorithm to the addition of one memory
element and one logic gate per bit, with respect to the
underlying iBDD.

Let us consider the decoding of the $i$th row component code,
in particular, the result of iBDD-SR corresponding to code bit
$c_{i,j}$. Let $\bar{\mu}_{i,j}^{\mathsf r, (\ell)} \in \{\pm1,
0 \}$ be the result of iBDD on code bit $c_{i,j}$ in iteration
$\ell$, where $0$ represents a decoding failure. In iBDD-SR,
the message sent to the $j$th column codes through $c_{i,j}$ is
computed as $\psi_{i,j}^{\mathsf{r},(\ell)}=\BB(\w_\ell \cdot
\bar{\mu}_{i,j}^{\mathsf r, (\ell)} + L_{i,j}),$ where
$w_\ell>0$ is a scaling parameter that can be optimized to
reduce the bit error rate (BER), $L_{i,j}$ is the reliability
of $c_{i,j}$, and $\BB(\cdot)$ takes the sign of the input and
maps $- 1 \mapsto 1$ and $+ 1 \mapsto 0$
(see~\cite{sheikh+:ECOC18} for details).

The range of $w_\ell$ is small and can thus be fixed to a
single value at the expense of a minor degradation in BER
performance~\cite{sheikh+:ECOC18}. In this case,
$\psi_{i,j}^{\mathsf{r},(\ell)}=\BB(\w \cdot
\bar{\mu}_{i,j}^{\mathsf r, (\ell)} + L_{i,j})$ can be
simplified to a comparison that returns
$\BB(\bar{\mu}_{i,j}^{\mathsf r, (\ell)})$ if $L_{i,j}<\w$ and
$\BB(L_{i,j})$ otherwise. Thus, we need only to store whether a
certain received bit has a lower reliability than $w$ and only
apply corrections for those bits, while rejecting attempts to
flip high-reliability bits; as we will show, this can be
implemented using low-complexity circuits. After performing a
given number of iterations using iBDD-SR, we perform two iBDD
iterations to correct erroneous bits mistakenly labeled as
reliable and thus not corrected during the iBDD-SR iterations,
cleaning up the few remaining errors.

We consider transmission over a binary-input additive white
Gaussian noise (BI-AWGN) channel, and a fixed scaling parameter
of 0.587. The output of the channel is quantized (without
$\EbNo$ normalization) to two bits: one HD bit and one reliability
bit. The decoder stores the channel output in two memories: one
reliability memory, which is only updated once a new block is
received, and one data memory, on which the component decoders
iteratively operate.

\section{VLSI Decoder Architecture}
\begin{figure}[t!]
  \centering
  \captionsetup[subfigure]{justification=centering}
  \begin{subfigure}[b]{0.45\textwidth}
  \includegraphics[width=\textwidth]{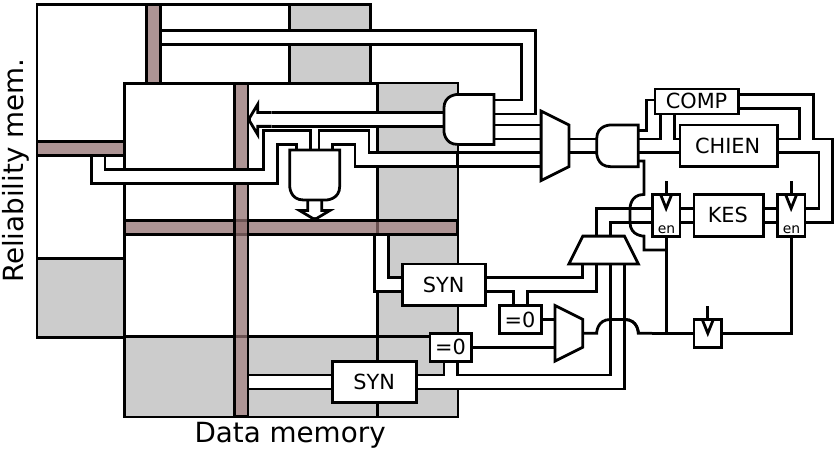}
  \caption{Block diagram of the soft-assisted product decoder.}
  \label{fig:bld}
  \end{subfigure}
	\hspace{2mm}
  \begin{subfigure}[b]{0.48\textwidth}
  \includegraphics[width=\textwidth]{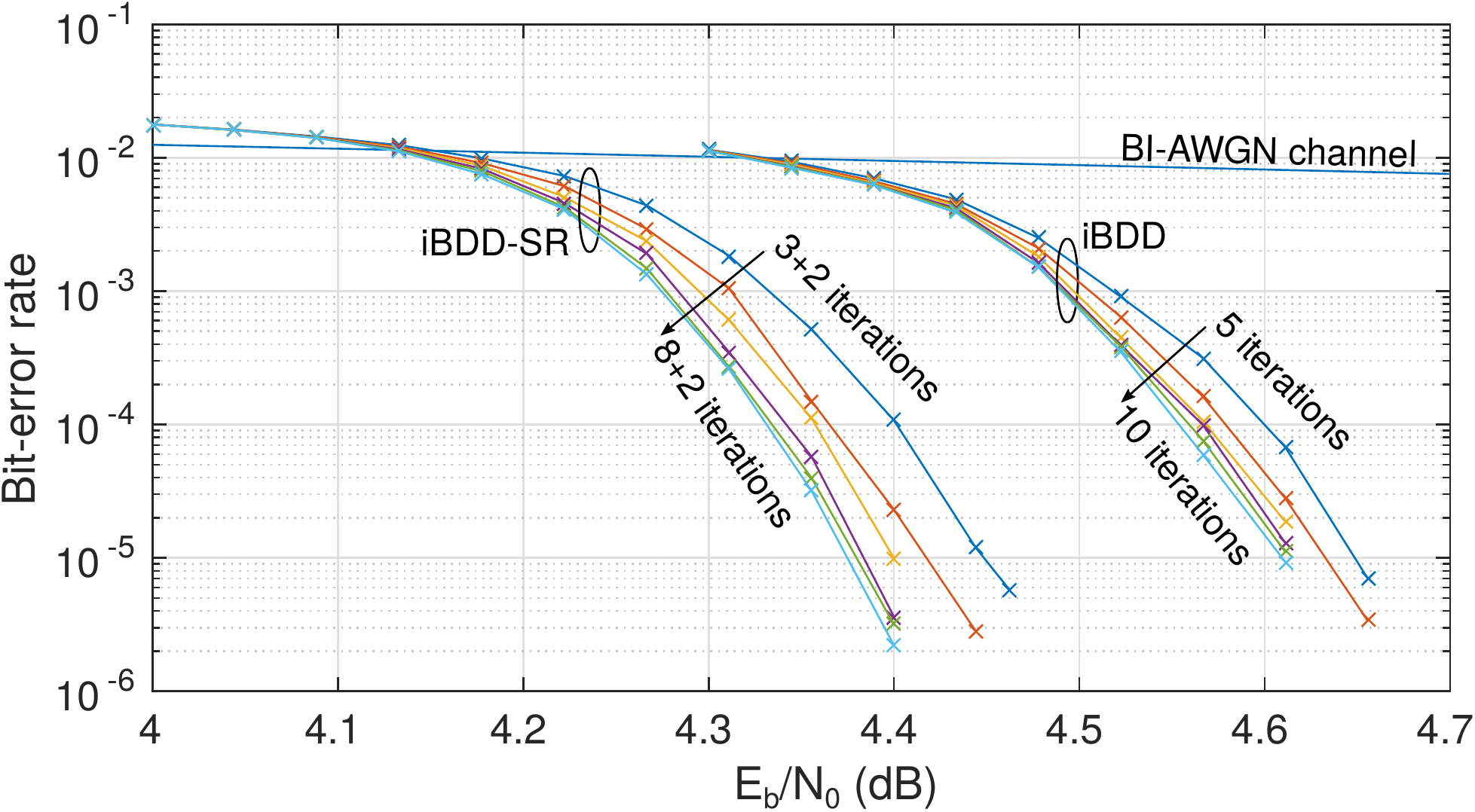}
         \caption{Output BER as a function of $\EbNo$.}
  \label{fig:ber}
  \end{subfigure}
  \vspace{2mm}
  \caption{Decoder circuit architecture and BER performance.}
  \vspace{-1.5mm}
\end{figure}

We consider a product code with BCH(255,231,3) component codes
giving an overall overhead of 21.9\% and a block length of
65,025 bits. The HD product decoder core, which shares
architectural features with the staircase decoder
in~\cite{fougstedt+:OFC2018}, employs fast and non-iterative
component decoders. This allows for very high throughputs,
while still keeping the vast majority of the logic circuits
static over row/column iterations, resulting in very low energy
dissipation. In addition to the HD bit, one reliability bit
has to be stored. The reliability bit is used to mask
corrections to bits which are considered reliable. This reduces
the miscorrection rate, while increasing the complexity by one
memory element and one AND-gate per bit.

Fig.~\ref{fig:bld} shows a block diagram of the implemented
decoder. The addition of circuits to handle reliability
information increases the total circuit area by a mere 5\%.
Since component decoders perform correction relatively
sporadically, we can employ extensive clock gating to reduce
the overall power dissipation. The reliability memory is
updated on the arrival of a new data block and kept idle
otherwise. Replicated syndrome-computation units are employed,
allowing us to keep their inputs largely static during iterations
and thus decreasing power dissipation at a small increase in
area. The decoder processing memory is only clocked at the
input of a new block and when a component decoder applies a
correction. The component decoder pipelines are sequentially
gated if a zero-syndrome is found.

\subsection{Decoder Performance Evaluation}

The hardware description (VHDL) implementations of the
soft-assisted (iBDD-SR) and HD (iBDD) decoders were simulated
using a VHDL-based quantized-output BI-AWGN channel. By
simulating the actual VHDL implementation, we verify the actual
decoder BER performance. Fig.~\ref{fig:ber} shows the output BER as a
function of $\EbNo$. Simulation run-time limits low-BER
statistics and we need to resort to extrapolation to estimate
the NCG at an output BER of $10^{-15}$. Here, we use \texttt{berfit} in
MATLAB to extrapolate to $10^{-15}$. Since low-BER statistics
are limited, we remark that the extrapolated results should be
seen as \emph{approximate} NCG. We estimate the NCG of iBDD-SR to 10.3--10.4\,dB and 10.1--10.2\,dB for
iBDD, as we vary the iteration count from 5 to 10. Thus, using
a single bit of soft information allows us to approach the NCGs
achieved by staircase codes at similar code
rates~\cite{zhang+:JLT14,fougstedt+:OFC2018}.

\subsection{Circuit Implementation and Evaluation}

The decoders were synthesized using Cadence Genus and a 0.9-V
28-nm fully-depleted silicon-on-insulator CMOS standard-cell
flow, with regular threshold voltage cells, characterized at the slow
process corner at 125$^\circ$C, using physical wire models and
a clock rate of 600\,MHz. The implemented decoders were
simulated using uniformly-distributed encoded data transmitted
over the implemented BI-AWGN channel model in order to generate
internal switching activity statistics for accurate power
dissipation estimation. The switching activity was then
back-annotated to the netlist in Cadence Genus, and power
dissipation estimation was performed at the typical process
corner and 25$^\circ$C.

\newpage

\section{Results and Discussion}

Table~\ref{table} shows the evaluation results of the
implemented product decoders. Decoder energy dissipation
depends on $\EbNo$ and, thus, the decoders are evaluated both
at $\EbNo=5.2$\,dB (corresponding to an input $\mathrm{BER}$ of
$10^{-2}$), and at the approximate $10^{-15}$ threshold (4.6 and
4.9\,dB for iBDD-SR and iBDD, respectively). Since the HD core
is common for both implementations, both the soft-assisted and
the HD decoder achieve up to 1\,Tb/s of information throughput.
However, the soft-assisted decoder achieves 0.2\,dB higher NCG
at the expense of just 5\% increase in area and gate count
(defined as cell area normalized to the smallest 2-input NAND
gate available in the standard-cell library), and
0.12--0.30\,pJ/bit increase in energy dissipation.
While the power dissipation decreases as more iterations are performed, the number of errors (and thus required component-code corrections) progressively reduces, and the energy per information bit increases since the decoder throughput is reduced.
The decoders
achieve a block-decoding latency of 53--103\,ns, depending on
the number of decoding iterations performed.

Decoder circuit implementations that achieve a throughput in
excess of 500\,Gb/s are rarely published in the open
literature. Compared to a recently published high-throughput SD
LDPC decoder~\cite{ghanaatian+:TVLSI2018}, our soft-assisted
decoder achieves 70\% higher throughput at less than half of
the area and a 97\% reduction in power dissipation, all while
achieving a higher NCG ($\EbNo=4.5$\,dB for a BER of $10^{-7}$,
compared to $4.97$\,dB in~\cite{ghanaatian+:TVLSI2018}). It
should be noted that the implementation
in~\cite{ghanaatian+:TVLSI2018} uses a short block-length
LDPC(2048, 1723) code. In comparison to our earlier HD staircase
decoders~\cite{fougstedt+:OFC2018}, the soft-assisted decoder
achieves 1\,Tb/s throughput with an area and energy dissipation
which is less than half of the HD staircase decoders, at similar
estimated NCGs.

\begin{table}[t!]
\centering
 \caption{VLSI Implementation Evaluation Results\label{table}}
 \renewcommand{\arraystretch}{1.05}
 \setlength{\tabcolsep}{5pt}
	\small
    \begin{tabular}{l||c|c|c|c|c|c||c|c|c|c|c|c} \toprule
    \label{tbl:numbent}
	    {\bf }&  \multicolumn{6}{c||}{{\bf Soft-Assisted Decoder (iBDD-SR)}}&  \multicolumn{6}{c}{{\bf Hard-Decision Decoder (iBDD)}}\\
    \midrule
	    {\bf Estimated net coding gain}&  \multicolumn{6}{c||}{{10.3 dB---10.4 dB}}&  \multicolumn{6}{c}{{10.1\,dB---10.2\,dB}}\\
              \bf{Cell area and gate count}        &\multicolumn{6}{c||}{4.76 mm$^2$, 9722\,kGates}&\multicolumn{6}{c}{4.52 mm$^2$, 9232\,kGates}\\
    \midrule
               \bf{Iterations}
		      &5&6&7&8&9&10&5&6&7&8&9&10\\
                   \midrule
	       \bf{Information throughput (Gb/s)}        &1000&842&728&640&572&516&1000&842&728&640&572&516\\
	       \bf{Power dissipation (mW) @5.2\,dB}&633&591&561&534&516&502&508&447&403&366&340&321\\

              \bf{Energy/info. bit (pJ/bit) @5.2\,dB}&0.63&0.70&0.77&0.83&0.90&0.97&0.51&0.53&0.55&0.57&0.59&0.62\\
\bf{Energy/info. bit (pJ/bit) @threshold}&0.81&0.88&0.95&1.01&1.08&1.16&0.61&0.63&0.65&0.66&0.69&0.72\\

			      \bf{Block decoding latency (ns)}&53&63&73&83&93&103&53&63&73&83&93&103\\
 \bottomrule
    \end{tabular}
\end{table}

\section{Conclusion}

We presented a soft-assisted product decoder implemented in a 28-nm process technology. The implemented iBDD-SR decoder relies on a high-throughput hard-decision decoder VLSI architecture and can achieve a throughput of 1\,Tb/s, while dissipating 0.63\,pJ/information bit. A coding gain of 10.3--10.4\,dB is achieved by exploiting one added soft-information bit and a 21.9\%-overhead code. Using soft information, if available, allows us to achieve similar coding gains as HD staircase decoders, with significantly lower circuit area and energy dissipation.

\vspace{1.5mm}

\noindent \textit{Acknowledgement: This work was financially supported by the Knut and Alice Wallenberg Foundation.}

\bibliographystyle{osajnl}

\end{document}